\documentclass[%
 aip,
 jmp,%
 amsmath,amssymb,
 reprint ]{revtex4-1}

\usepackage{graphicx}
\usepackage{color}
\usepackage{dcolumn}
\usepackage{bm}

\begin{document}

\preprint{AIP/123-QED}

\title[Analysing international events through the lens of statistical physics: the case of Ukraine]{Analysing international events through the lens of statistical physics: the case of Ukraine}

\author{M. Zanin}
 \email{massimiliano.zanin@gmail.com.}

\author{J. H. Mart\'inez}
\affiliation{ 
Instituto de F\'isica Interdisciplinar y Sistemas Complejos IFISC (CSIC-UIB), Campus UIB, 07122 Palma de Mallorca, Spain
}%

\date{\today}

\begin{abstract}
During the last years, statistical physics has received an increasing attention as a framework for the analysis of real complex systems; yet, this is less clear in the case of international political events, partly due to the complexity in securing relevant quantitative data on them. Here we analyse a detailed data set of violent events that took place in Ukraine since January 2021, and analyse their temporal and spatial correlations through entropy and complexity metrics, and functional networks. Results depict a complex scenario, with events appearing in a non-random fashion, but with eastern-most regions functionally disconnected from the remainder of the country - something opposing the widespread ``two Ukraines'' view. We further draw some lessons and venues for future analyses.
\end{abstract}


\maketitle

\begin{quotation}
Statistical physics is becoming a reference framework for studying real-world complex systems, thanks to the possibilities it offers to connect micro-scale dynamical properties with macro-scale observations. In spite of this, political events have mostly been neglected, partly due to the complexity in securing relevant quantitative data on them - the solution usually being relying on indirect data, as e.g. Twitter activity. We here leverage on a detailed data set of violent events that preceded the current military crisis in Ukraine, and use two concepts (entropy and complexity, and functional networks) to unveil the underlying relationship structure.
\end{quotation}

\section{\label{sec:Intro}Introduction}

During the last decades, statistical physics concepts and tools have ceased being exclusive of this scientific field, for becoming standard approaches used in the analysis of numerous and heterogeneous real-world problems. To illustrate but a few examples, complex networks have become an essential asset in epidemics spreading models \cite{kiss2017mathematics}, neuroscience \cite{bullmore2009complex}, or climate \cite{donges2009complex, kretschmer2016using}; and entropy and irreversibility have been used to characterise biomedical systems, from brain \cite{inouye1991quantification, abasolo2006entropy, echegoyen2020permutation} to heart dynamics \cite{costa2008multiscale}. The reason for such success is possibly rooted in statistical physics' ability for decoupling the dynamical and observational scales; that is, a system may only be observable at the macro-scale, but conclusions about the underlying micro-scale dynamics can still be drawn. 

Among the real-world problems that have still received little to no benefit from statistical physics concepts, the analysis of international (violent) events stands out. This may be due to several reasons, from the difficulty in securing quantitative data on those events (which usually are of indirect nature, e.g. Twitter messages \cite{makhortykh2015savedonbasspeople, makhortykh2017social}), to the natural barriers to the cross-dissemination between social and physical sciences.
The objective of this contribution is to bridge this gap, and specifically to showcase how statistical physics concepts (namely, entropy, complexity, and functional networks) can be used to improve our understanding of international events. We specifically focus on the Ukrainian crises, and its role in the ongoing conflict between this country and the Russian Federation.

Ukrainian crises can partly be seen as an old process, resulting the juxtaposition of two national identities, i.e. the pro-European west part of the country and the pro-Russian east, which have failed to coexist peacefully \cite{zhurzhenko2014divided, gotz2016russia, kuzio2018russia, harris2020role}. A turning point can be found in November 2013, when protests erupted against Ukrainian President Viktor Yanukovych's decision to reject a stronger economic integration deal with the European Union. This resulted, on one hand, in Russian military troops taking control of the region of Crimea in March 2014, which was later annexed by the Russian Federation following a local referendum \cite{grant2015annexation}. This initial military conflict then escalated in a full war in February 24th 2022, with the Russian Federation launching a full-scale military invasion into Ukraine - still developing at the time when this manuscript was being prepared. Beyond these large-scale military operations, many local violent events took place since 2014, and especially during the last year. These involved Ukrainian security forces, pro-Russian anti-government separatist groups, and the general population; and included from bombardment and shelling, to unrests and protests.

In this contribution we analyse a public data set of violent events happened in Ukraine from January 2021 to January 2022, by applying two complementary approaches: entropy and complexity on one hand, and functional networks on the other. These are respectively aimed at detecting temporal and spatial relationships in the appearance of those events. In other words, we try to answer the question: are events independent from each other, both in a temporal and spatial dimensions? Or, on the other hand, do past events in one region affect the appearance of other events? Results indicate that the situation is more similar to the latter case, with events having both temporal and spatial structures not compatible with a random dynamics. Most interestingly, events in the regions of Donetsk and Luhansk, i.e. the two regions contested between Ukraine and the Russian Federation, are causally disconnected from the remainder of the country. The implications of these results are discussed, and we finally draw some conclusions and lessons learnt.

\section{\label{sec:DataAndMethods}Data set and analysis techniques}

\begin{figure*}
\includegraphics[width=0.9\textwidth]{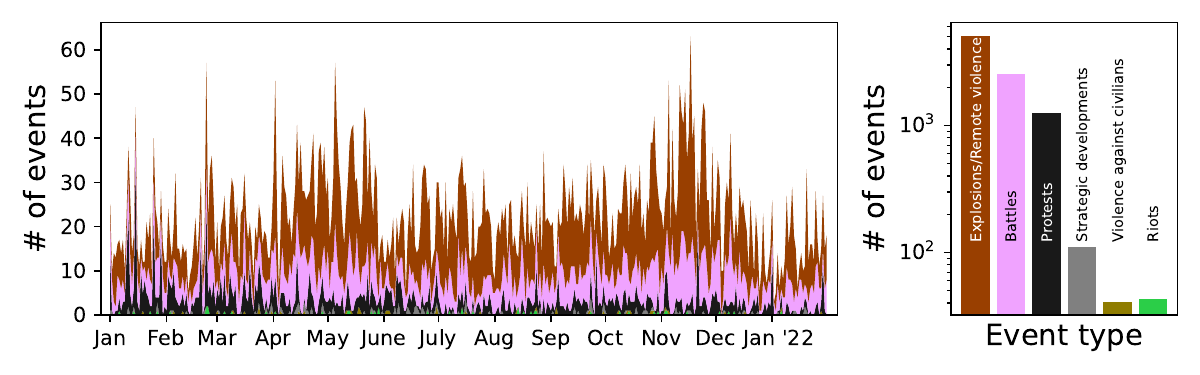}
\caption{\label{fig:AllEvents} Violent events in Ukraine. The figure depicts the temporal evolution of the number of events by day, according to their categorisation into six groups, left panel (see right panel for colour code); and the total number of events by group, right panel. }
\end{figure*}

Data about unrest and other violent events in Ukraine have been obtained from the Armed Conflict Location \& Event Data Project (ACLED)~\cite{acled2010introducing}, and are freely available at \url{https://acleddata.com}. They contain a list of all events by date, including the parties or group in them involved, the type of the event, and a geolocalisation. Regarding the latter, and in order to avoid a too granular division of data, only the first administrative division (regions, or {\it oblasts}) has here been considered. A total of $9,104$ events are reported from January 1st 2021 to January 31st 2022; the evolution of the number of events and their type is reported in Fig. \ref{fig:AllEvents}.

This data set has been analysed through two complementary statistical physics techniques: permutation patterns and functional complex networks. For the sake of completeness, these are briefly described below.

\subsection{Permutation entropy and complexity}
\label{sec:complexity}

Permutation entropy and complexity are named after the symbols, i.e. the so-called permutation patterns, over whose probability distribution function (PDF) the two metrics are computed. The method to retrieve those patterns translates the relative amplitudes of a time series in observational windows of a certain length $D$ into their symbolic representation~\cite{bandt2002permutation}, thus accounting for the temporal causality in the data. This allows to assess dynamical properties like degree of randomness or periodicity, possible degrees of stochasticity, or level of complexity \cite{rosso2007distinguishing, zanin2021ordinal}. This procedure is known to be simple, fully data-driven, computationally efficient, robust against noise, and useful for data with weak stationarity \cite{amigo2010permutation}.

Given a time series $X_t \mid \{x_t \in \mathbb{R}\}\:\forall \: t=1,\dots M$, we define a pattern length of $D = 4$, and divide the time series into $T=M-(D-1)$ overlapping windows. The elements of each $D$-window $X_s = (x_s, x_{s+1},\dots,x_{s+(D-1)})$ of $X_t$ are sorted in increasing order to capture their indexes $i_0, i_1,\dots,i_{(D-1)}$, such that $x_{s+i_0}\leq x_{s+i_1}\leq \dots \leq x_{s+i_{(D-1)}}$. Hence words $\pi = (i_0, i_1,\dots,i_{(D-1)})$ are symbols representing all possible $D!$ permutations of $\{0,1,2,\dots,D-1\}$, this way encoding the relative amplitudes of each $D$-segment. The associated PDF $p(\pi)$ associates the occurrence frequency of permutation patterns $\pi$ in the data, satisfying $\sum_{\pi}p(\pi)=1$. 

We describe the temporal dynamics of violent events through both the their permutation entropy ($H$) and complexity ($C$). The former is a normalised quantifier that captures correlation structures not potentially retrieved by a simple entropy analysis, and is defined as:

\begin{equation}
H[p] = \frac{-1}{S_{max}}\sum_{\pi=1}^{N}p(\pi)log(p(\pi)).
\end{equation}

$H$ measures the order level of the system, and is given by the ratio of the entropy of $p(\pi)$ to the maximum entropy of the system modelled by a uniform probability $p_e \mid S_{max} = log(D!)$. Therefore $0\leq H[p]\leq 1$, for a total ordered, or uncorrelated random system, respectively. 

In a complementary way, $C$ characterises the system organisation accounting for the interplay between its order and disorder, and is independent of size effects since it does not increase when the system becomes larger. It is defined as the product of $H$, and a non-euclidean similarity between the observed and $p_e$: $C[p] = H[p]\cdot Q[p,p_e]$, where $Q[p,p_e]$ is the Jensen-Shannon divergence. It assesses the emergence of correlation structures, with $C=0$, in the case of order ($H=0$) or total randomness ($H=1$), implying no structure in the time series.

Statistical significance of results is assessed through surrogate time series. Specifically, we compute $H$ (respectively, $C$), for then getting the $H_r$ ($C_r$) counterpart in a set of $1,000$ of randomly shuffled time series. We firstly analyse the global structure of the aggregated violent events during a year in the whole country ($M=396$), for then focusing on the time evolution for windows of $150$ days. 
This window size has been chosen as a compromise between having too large, which would smooth out any fast dynamics, and too short time windows, which would hinder the statistical significance. For reference, results for $75$ and $300$ days are also reported.
Note that the embedding dimension has here been set to $D = 4$; while other options could be explored, this value has been chosen to be the largest fulfilling the condition $150 > (D+1)!$, hence being large enough to capture complex relations in the data, but also small enough to ensure statistical significance of results in the $150$-days windows \cite{amigo2007true, riedl2013practical}.

\subsection{Functional complex networks}

We further analyse how events are connected in the spatial dimension by leveraging on the concept of functional networks. This approach entails reconstructing complex network representations \cite{strogatz2001exploring, albert2002statistical, newman2003structure, boccaletti2006complex} of a system, based on detecting relationships between its constituting elements through the analysis of their temporal dynamics. It has received a special attention from the neuroscience community, in which it has allowed to unveil the patterns of interactions between brain regions in health and pathologies \cite{bullmore2009complex, park2013structural, martinez2018role, korhonen2021principles}; but it has also been applied to e.g. climate modelling \cite{donges2009complex, kretschmer2016using}, ecology \cite{messier2019functional} or air transport management \cite{pastorino2021air}. In the context of this work, nodes of the network represent Ukrainian regions, pairwise connected whenever a statistically significant relationship between the corresponding time series of events is detected.

Relationships are here detected through the celebrated Granger causality test \cite{granger1969investigating}, developed by the economy Nobel Prize laureate Clive Granger on top of the prediction theory of Norbert Wiener \cite{wiener1956theory}, and one of the most well-known statistical tests for evaluating the presence of {\it predictive causality} \cite{diebold1998elements} between pairs of time series. Note that, while the test name includes the word causality, it does not necessarily measure true causality \cite{granger1988causality}; it instead quantifies the information transfer across multiple time scales. In spite of this, and for the sake of simplicity, the relationships detected by this test will here be called {\it causal}. It is additionally worth noting that other causality tests have been proposed in the Literature \cite{staniek2008symbolic, pearl2009causality, peters2016causal}, although their applicability of size-limited time series is not always straightforward.

A brief description of the test is here included. Let us consider two elements ${A}$ and ${B}$, respectively described by two time series $a$ and $b$. Two autoregressive-moving-average (ARMA) models are fitted on the data, respectively called the restricted and unrestricted regression models:

\begin{eqnarray}
    \label{eq:gc}
	a_t = K \cdot a_{t-1}^m + \epsilon_t, \\
	a_t = K' \cdot \left( a_{t-1}^m \oplus b_{t-1}^m \right) + \epsilon'_t.	
\end{eqnarray} 

$m$ here refers to the model order, the symbol $\oplus$ denotes concatenation of column vectors, $K$ and $K'$ contain the model coefficients, and $\epsilon_t$ and $\epsilon'_t$ are the residuals of the models. A Granger causality is then detected if $\sigma^2( \epsilon'_t ) < \sigma^2( \epsilon_t )$, i.e. if including past information of the driving time series helps predicting the future of the driven one. In order to assess the statistical significance and obtain a $p$-value, an F-test is performed to check whether the coefficients $K'$ associated to the time series $b$ are different from zero - i.e. whether $b$ is actually having an impact in the prediction.

As can be seen from Fig. \ref{fig:AllEvents}, the time series here considered are not stationary, as a higher number of events are present around May and December 2021. In order to solve this, the Granger test has been applied to normalised time series, representing the fraction of events observed in one day in each Ukrainian region over the total number of events in the same day. The final results are networks composed of $26$ nodes, one for each region; and are represented by adjacency matrices $A$, of size $26 \times 26$, where the element $a_{ij}$ has a value of $1$ to indicate that there is a directed edge from node $i$ to $j$ (i.e. the events in region $i$ ``Granger-cause'' events in region $j$), and $0$ otherwise \cite{albert2002statistical, boccaletti2006complex}. In order to avoid the increased probability of type I errors as a consequence of the multiple comparisons required by the reconstruction process, we applied a Bonferroni correction and rejected the null hypothesis of the test for an effective $\alpha = 0.01 / (26 \times 25) \approx 1.53 \cdot 10^{-5}$. Additionally, the significance of the degree of the most connected nodes is tested using networks created with randomly shuffled time series - i.e. time series in which the temporal structure is destroyed.

As a final note, and in a way similar to what described in Sec. \ref{sec:complexity}, functional networks have also been reconstructed for rolling windows of $150$ days, to explore the evolution of causal relationships.

\section{Results: temporal relationships}
\label{sec:temporal}

We start by analysing the temporal relationships in the data, in order to understand if events have appeared randomly in the considered time window, or if instead they present some kind of internal structure. To this end, the two metrics described in Sec. \ref{sec:complexity} have been calculated over the time series representing the total number of events per day. The results, reported in the top row of Tab. \ref{tab:entropy}, indicate that this time series is highly irregular, but yet with some non-random structure (note the Z-Score close to $\pm 2$). The same table also reports the values for the regions of Crimea, Donetsk and Luhansk, i.e. the three regions that have mostly been contested between Ukraine and the Russian Federation; these results will be used in the next section.

\begin{figure}
\includegraphics[width=0.48\textwidth]{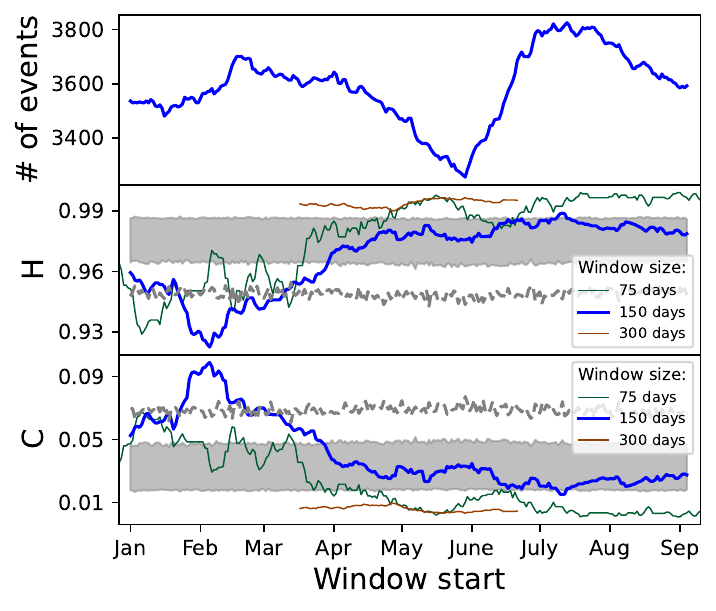}
\caption{\label{fig:HC} Statistical properties of the time series of violent events. The three panels, from top to bottom, report for a $150$-days rolling window: (i) the evolution of the total number of events in the data set; (ii) the entropy $H$; and (iii) the statistical complexity $C$. Grey bands in the central and bottom panels correspond to the $10-90$ percentile of values obtained through a random shuffling of the data, and the dashed grey line the $99$ percentile. Additionally, the green and brown lines correspond to the result using rolling windows of respectively $75$ and $300$ days, aligned according to the central day in each window.}
\end{figure}

Fig. \ref{fig:HC} reports the evolution of $H$ (middle panel) and $C$ (bottom panel) for rolling windows of $150$ days, along with the evolution of the corresponding number of events for reference (top panel). Additionally, the grey bands indicate the $10$-$90$ percentiles of the same metrics calculated over randomly shuffled versions of the same time series; and the dotted grey lines the $99$ percentile. It can be observed that the time series are mostly compatible with a random noise, except for windows starting around February 2021. For reference, the same results are reported for rolling windows of $75$ (green lines) and $300$ (brown lines) days, showing in the former case a similar behaviour.

\begin{table*}
\caption{\label{tab:entropy}Entropy $H$ and complexity $C$ (Z-Score in parenthesis) for the time series of the number of total events in Ukraine, and for the regions of Crimea, Donetsk and Luhansk. The third and fifth columns also report the average and standard deviation of the two metrics, calculated on $10^3$ randomly shuffled versions of the same time series.}
\label{table}
\setlength{\tabcolsep}{3pt}
\center
\begin{tabular}{|l|l|l|l|l|}
\hline
{\bf Region} & {\bf H (Z-Score)} & {\bf Rnd H} & {\bf C (Z-Score)} & {\bf Rnd C} \\ \hline
Ukraine & $0.9839$ ($-1.94$) & $0.9907 \pm 0.0035$ & $0.0209$ ($1.87$) & $0.0122 \pm 0.0046$ \\ \hline
Crimea & $0.3183$ ($-3.37$) & $0.3348 \pm 0.0049$ & $0.2318$ ($-1.34$) & $0.2384 \pm 0.0049$ \\ \hline
Donetsk & $0.9741$ ($-3.21$) & $0.9898 \pm 0.0049$ & $0.0326$ ($3.91$) & $0.0134 \pm 0.0049$ \\ \hline
Luhansk & $0.9740$ ($-1.39$) & $0.9841 \pm 0.0072$ & $0.0326$ ($1.62$) & $0.0208 \pm 0.0072$ \\ \hline
\end{tabular}
\end{table*}

\section{Results: spatial relationships}
\label{sec:spatial}

\begin{figure}
\includegraphics[width=0.48\textwidth]{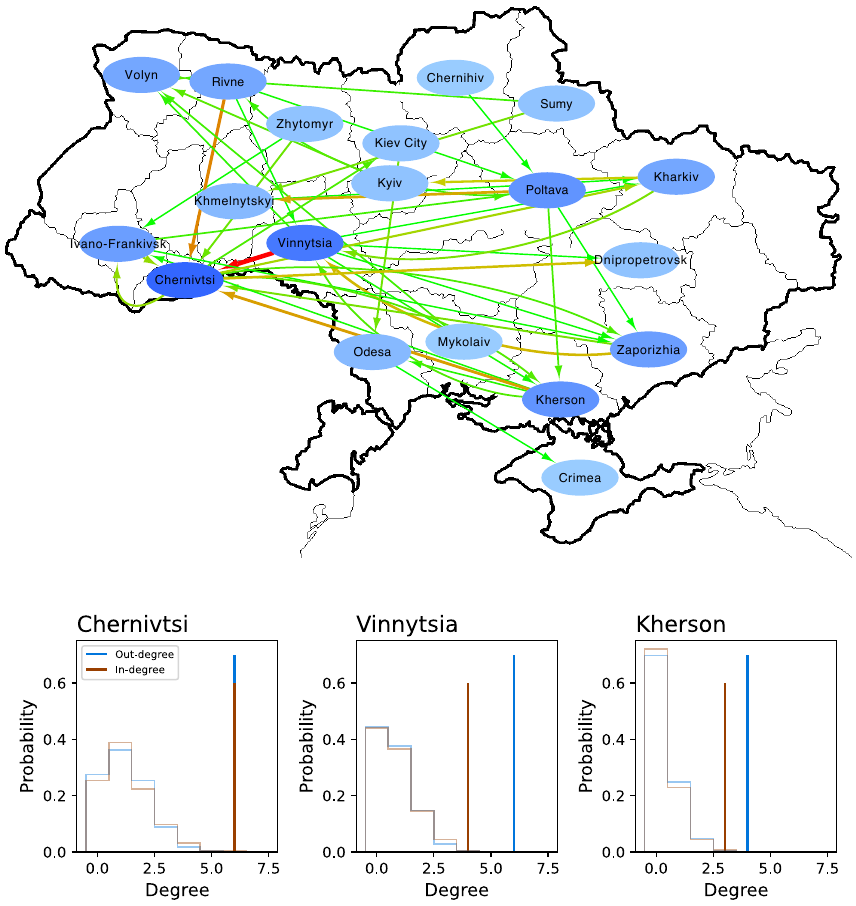}
\caption{\label{fig:Network} Graphical representation of the Ukrainian functional network - regions not connected to the network are not represented. Link colours, from green to red, represent the strength of the functional connection (inversely proportional to the corresponding $p$-value); node colours, from light to dark blue, the degree of nodes. The bottom panel reports the out- and in-degrees of the most connected nodes. Light lines report the distribution of the degrees of the same node, obtained with $10^3$ random shuffling of the time series.}
\end{figure}

Complementary to what presented in Sec. \ref{sec:temporal}, we here analyse the relationships between events happened in different regions of Ukraine using the functional network approach. Specifically, Fig. \ref{fig:Network} depicts all links between pairs of Ukranian regions that are statistically significant, as yielded by applying the Granger Causality test over the time series of events in each Ukrainian region. Two main facts stand out. Firstly, the network is relatively highly connected (link density of $0.063$), much more that what would be expected if events were random ($0.024 \pm 0.0059$). Similarly to what observed in Sec. \ref{sec:temporal}, events do not appear independently, but are instead causally connected also in the spatial dimension, suggesting a kind of action-reaction mechanism. Secondly, links are spread across the whole country, but interestingly the two most eastern regions (Donetsk and Luhansk, i.e. the two regions contested between Ukraine and the Russian Federation) are not connected to the network. This is not due to a reduced number of violent events, respectively of $5,757$ and $1,988$. It is also not due to a lack of temporal structure in the events of those two regions, as illustrated by the results in Tab. \ref{tab:entropy} - note the large Z-Score, especially for the region of Donetsk. It can therefore be concluded that violent acts in those two regions are not the cause, or the result, of events happening in the remainder of the country, but that instead have an independent dynamics.

The bottom part of Fig. \ref{fig:Network} reports the degree (both out-, blue lines, and in-, brown lines) of the three most connected nodes, and the corresponding distribution obtained with randomly shuffled time series. While degrees are generally small, also due to the reduced size of the network, they are statistically significant ($p$-value always smaller than $10^{-3}$). Most notably, the two most connected regions are Chernivtsi and Vinnytsia, located on the western part of the country - once again, not geographically connected with the contested regions of Donetsk and Luhansk.

\begin{figure}
\includegraphics[width=0.48\textwidth]{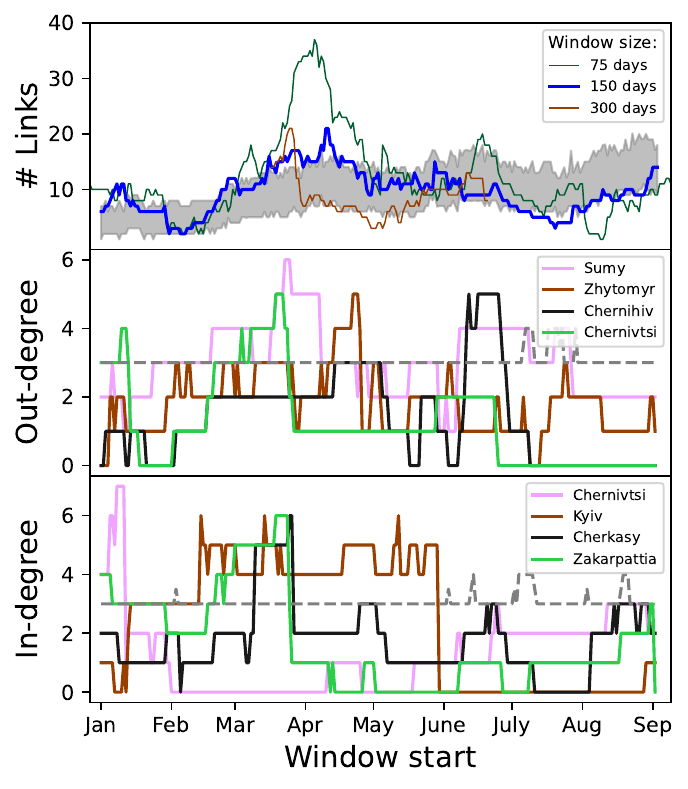}
\caption{\label{fig:DegreesEvol} Evolution of the connectivity for networks reconstructed with a rolling window of $150$ days. The top panel reports the evolution of the number of links (blue line) and of the $1$-$99$ percentile (grey band). The green and brown lines correspond to the result using rolling windows of respectively $75$ and $300$ days, aligned according to the central day in each window. The middle and bottom panels depict the evolution of the out- and in-degrees of the four regions that have reached the largest value; the grey dashed lines represent the maximum degree obtained in $10^3$ networks reconstructed by randomly shuffling the time series. }
\end{figure}

We have finally analysed how the connectivity has evolved through time. For that, once again a rolling window of $150$ days has been considered, and a functional network has been reconstructed for each time interval. The top panel of Fig. \ref{fig:DegreesEvol} reports the evolution of the number of links (blue line); for reference, the $1$-$99$ percentiles obtained in networks of randomly shuffled data are also reported - see the grey band. Additionally, as in Fig. \ref{fig:HC}, the same results are reported for rolling windows of $75$ (green lines) and $300$ (brown lines) days. It can be appreciated that the number of links is seldom statistically significant, most probably due to the reduced time series length, which makes the estimation of the Granger causality unreliable. The middle and bottom panels of the same figure further report the evolution of the out- and in-degrees of the four regions that have reached the largest connectivity; once again, they seldom are statistically significant (see the grey dashed lines, representing the maximum degree obtained in networks of randomly shuffled data).

\section{\label{sec:Concl}Discussion and conclusions}

In this contribution we have showcased how two statistical physics concepts, namely entropy and complex networks, can be used to extract information from violent international events, with a specific focus on the developing Ukrainian crisis. In spite of the conceptual simplicity of this analysis, and of some major limitations that are discussed below, some interesting conclusions can be drawn. Events are not independent, both in the temporal and spatial scales. This is not surprising, as some events are inherently answers to others, e.g. protests can be the response to other violent events; also, battles and explosions are usually part of larger-scale plans, and are therefore coordinated.

\begin{figure}
\includegraphics[width=0.48\textwidth]{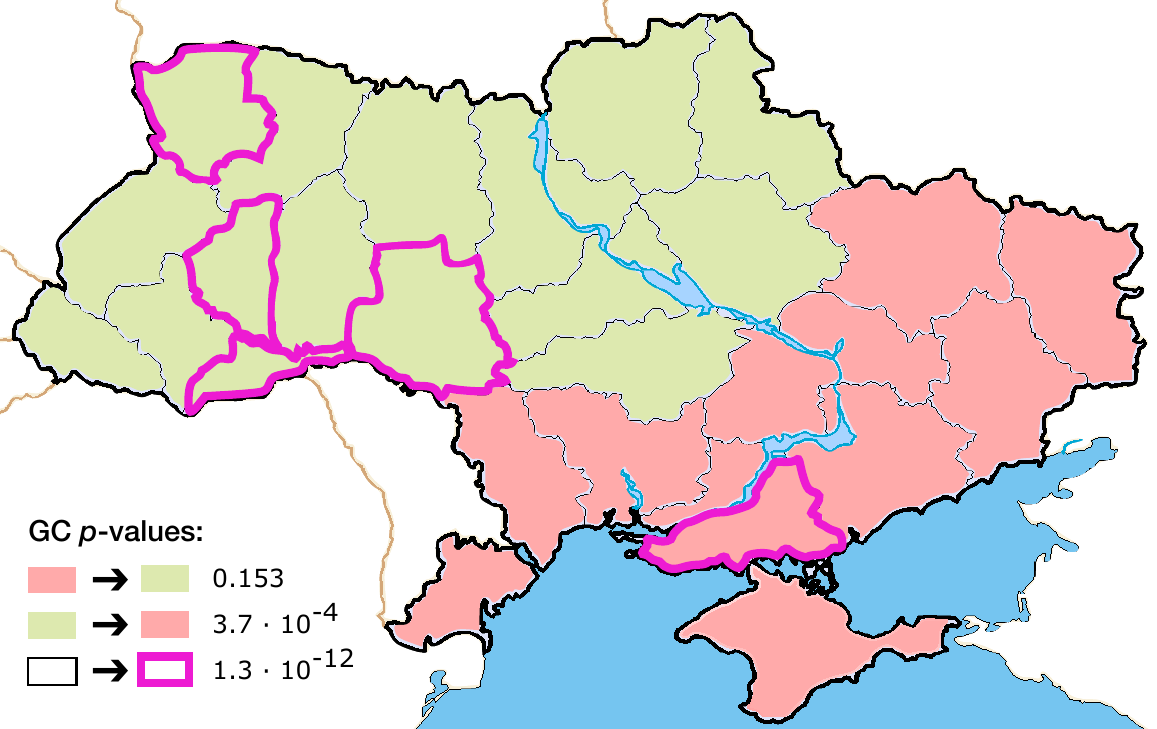}
\caption{\label{fig:TwoRegions} Testing the ``two Ukraines'' hypothesis. Light red and green provinces respectively correspond to the pro-Russian and pro-west regions, according to the results of the Ukrainian presidential elections of 2010. Provinces outlined in purple correspond to those minimising the $p$-value of the Granger causality test - values are reported in the bottom left part. }
\end{figure}

It is nevertheless interesting to see some unexpected patterns. Specifically, the regions of Donetsk and Luhansk are not causally connected (in the Granger sense) to any other region, and Crimea only receives a rather weak link. This seems to indicate that violent events in the country are independent, and not a consequence, of what there happens, in spite being these the three regions most contested between Ukraine and the Russian Federation - and actually being the target of the ongoing Russian armed invasion. This contradicts the standard ``two Ukraines'' view of a country divided between pro-west and pro-Russian regions, a view that has already received some criticisms \cite{zhurzhenko2014divided, palermo2020elephant}. This is supported by the fact that Granger-testing the time series representing the aggregated events in those two regions yield large $p$-values, of $0.153$ (east to west) and $\approx 10^{-4}$ (west to east) - see Fig. \ref{fig:TwoRegions}.
On the contrary, these results suggest a more complex situation, with tensions distributed throughout the whole country. This supports a view according to which the main divide in Ukraine is essentially a Ukrainian vs. Ukrainian one, with no sharp and unambiguous division along ethnic, geographical or even linguistic lines \cite{riabchuk2015two, frye2015voters}. We tested this by executing an optimisation problem, aimed at detecting the two sets of regions minimising the $p$-value of the Granger causality \cite{zanin2021simplifying}. The result, represented in Fig. \ref{fig:TwoRegions}, suggests that violent events in five provinces (Volyn, Ternopil, Chernivtsi, Vinnytsia and Kherson) are driven by events in the remainder of the country. Most importantly, this analysis is objective in nature, thus not relying on the interpretation of political events or of other qualitative types of data.

The analysis here presented also highlights some limitations, associated with the considered type of data. The relative low number of violent events observed in the country precludes the possibility of an analysis on smaller temporal and spatial scales, as this would otherwise require dealing with time series with many equal and/or zero values - a known limitation of the permutation entropy approach \cite{zunino2017permutation, cuesta2018patterns}. This is apparent even in the analysis here presented, with results for a $150$-days rolling window seldom being statistically significant. The magnitude of events, which may provide a more complete view of how the situation unfolded, has also been disregarded. Taking this aspect into consideration is nevertheless not trivial, as would require, firstly, extract quantitative information from the text describing each event, e.g. through Natural Language Processing \cite{nadkarni2011natural}; and secondly, designing a way of comparing heterogeneous magnitudes, e.g. number of people in a protest vs. number of deaths in a military attack.

\section*{Acknowledgments}

This project has received funding from the European Research Council (ERC) under the European Union's Horizon 2020 research and innovation programme (grant agreement No 851255). J.H.M. acknowledges funding from the project PACSS RTI2018-093732-B-C22 of the MCIN/AEI/10.13039/501100011033/, and by EU through FEDER funds (A way to make Europe). The authors acknowledge the Spanish State Research Agency through Grant MDM-2017-0711 funded by MCIN/AEI/10.13039/501100011033. The authors thank J. J. Ramasco for his assistance.

\bibliography{Ukraine}

\end{document}